\documentclass[amsfonts, amssymb, pre, superscriptaddress, notitlepage, twocolumn, 11pt, tightenlines, floatfix]{revtex4-2}

\usepackage{amsmath,url}
\usepackage{graphicx,sidecap}
\usepackage{CrimsonPro}
\usepackage[letterpaper, margin=20mm]{geometry}
\begin{document}

\title{Where postdoctoral journeys lead}

\author{Yueran Duan}
\affiliation{School of Economics and Management, China University of Geosciences, Beijing, China}
\affiliation{Department of Computer Science, Aalto University, Espoo, Finland}

\author{Shahan Ali Memon}
\affiliation{Information School, University of Washington, Seattle, USA}
\affiliation{Social Science Division, New York University Abu Dhabi, United Arab Emirates}

\author{Bedoor AlShebli}
\affiliation{Social Science Division, New York University Abu Dhabi, United Arab Emirates}

\author{Qing Duan}
\affiliation{School of Economics and Management, China University of Geosciences, Beijing, China}

\author{Petter Holme}
\affiliation{Department of Computer Science, Aalto University, Espoo, Finland}
\affiliation{Center for Computational Social Science, Kobe University, Kobe, Japan}

\author{Talal Rahwan}
\affiliation{Computer Science Program, Science Division, New York University Abu Dhabi, United Arab Emirates}

\begin{abstract}
Postdoctoral training is a career stage often described as a demanding and anxiety-laden time when many promising PhDs see their academic dreams slip away due to circumstances beyond their control. We use a unique data set of academic publishing and careers to chart the more or less successful postdoctoral paths. We build a measure of academic success on the citation patterns two to five years into a faculty career. Then, we monitor how students' postdoc positions---in terms of relocation, change of topic, and early well-cited papers---relate to their early-career success. One key finding is that the postdoc period seems more important than the doctoral training to achieve this form of success. This is especially interesting in light of the many studies of academic faculty hiring that link Ph.D.\ granting institutions and hires, omitting the postdoc stage. Another group of findings can be summarized as a Goldilocks principle: it seems beneficial to change one's direction, but not too much.
\end{abstract}

\maketitle

\section{Introduction}

The postdoc is the first career stage specialized for academia. In that sense, it is the first occasion where leaving academia could be seen as a failure. Yet, it is a bottleneck---many postdocs will be unsuccessful in finding faculty positions. In academia, postdocs are the ones with the most time on their hands to produce the core output of academia---science. Since postdocs have little to blame a meager output on, the stakes become sky-high. It is not surprising that the anecdotal picture of the postdoc experience is bleak---``disenchanted''~\cite{woolston2020postdoc}, ``stressed-out''~\cite{arnold2014stressed}, ``unhappy''~\cite{grinstein2017unhappy}, and ``exploited''~\cite{stephan2013exploit} are all recent titular epithets in the literature. However, not all postdocs leave academia. Some go on to become scientific top-achievers, so what commonalities do these successful cases share? 

The emerging interface between data science and the study of academic knowledge production---the \textit{science of science}~\cite{sos})---not only brings us a new understanding of the societal enterprise behind it, but also has the potential to rectify structural biases and troubling trends in academic science. For example, recent studies focusing on American academia have shown that the prestige of the Ph.D.\ awarding institution has a significant impact on not only faculty hiring~\cite{clauset_systematic}, but also retention and attrition of faculty members~\cite{wapman_quantifying}. Essentially, a few high-prestige universities educate faculty across the American academia. Subsequent studies have argued that the dynamics behind faculty hiring perpetuate these structural biases.

A vast majority of data-driven studies of the academic job market include the Ph.D.\ granting institution as an explanatory variable~\cite{clauset_systematic,wapman_quantifying,eun_dynamic,cowan_rossello,fernandes_etal,fowler_etal,hanneman}, but very few study the impact of the postdoc period. Two exceptions are works by Fernandes \textit{et al.}~\cite{fernandes_etal} and Horta \cite{horta2009}, finding that being awarded a postdoctoral fellowship is as vital for a successful early academic career as a top-tier first-author publication and that postdoc experience helps build long-lasting international networks. One possible reason for this lack of data-driven studies of postdocs is the prevailing descriptions of postdocs in passive terms---postdocs are a waiting or probation period~\cite{i_r,s_ma}. Papers usually take a systemic perspective---asking what the postdoctoral training contributes to society~\cite{aalund2020academic,i_r,alberts_addressing}---or focusing on the well-being of postdoctors~\cite{arnold2014stressed,grinstein2017unhappy,stephan2013exploit}.

With this paper, we aim to take a first step to fill the void of comprehensive, statistically grounded career advice for newly-minted doctors. To shed light on the postdoctoral bottleneck and the postdoctoral training's role in early professorship, we collect a unique dataset drawing information both from a publication database (Microsoft Academic Graph, MAG) and a large online professional network (see Materials and Methods). The data covers 45,572 careers, spanning 25 years, from all academic disciplines and all over the Earth. For simplicity, we exclude careers other than those involving a postdoc in between academia and careers entirely within or without academia.

\section*{The postdoc-to-faculty bottleneck}

\begin{figure*}[t]
\includegraphics[width=0.7\textwidth]{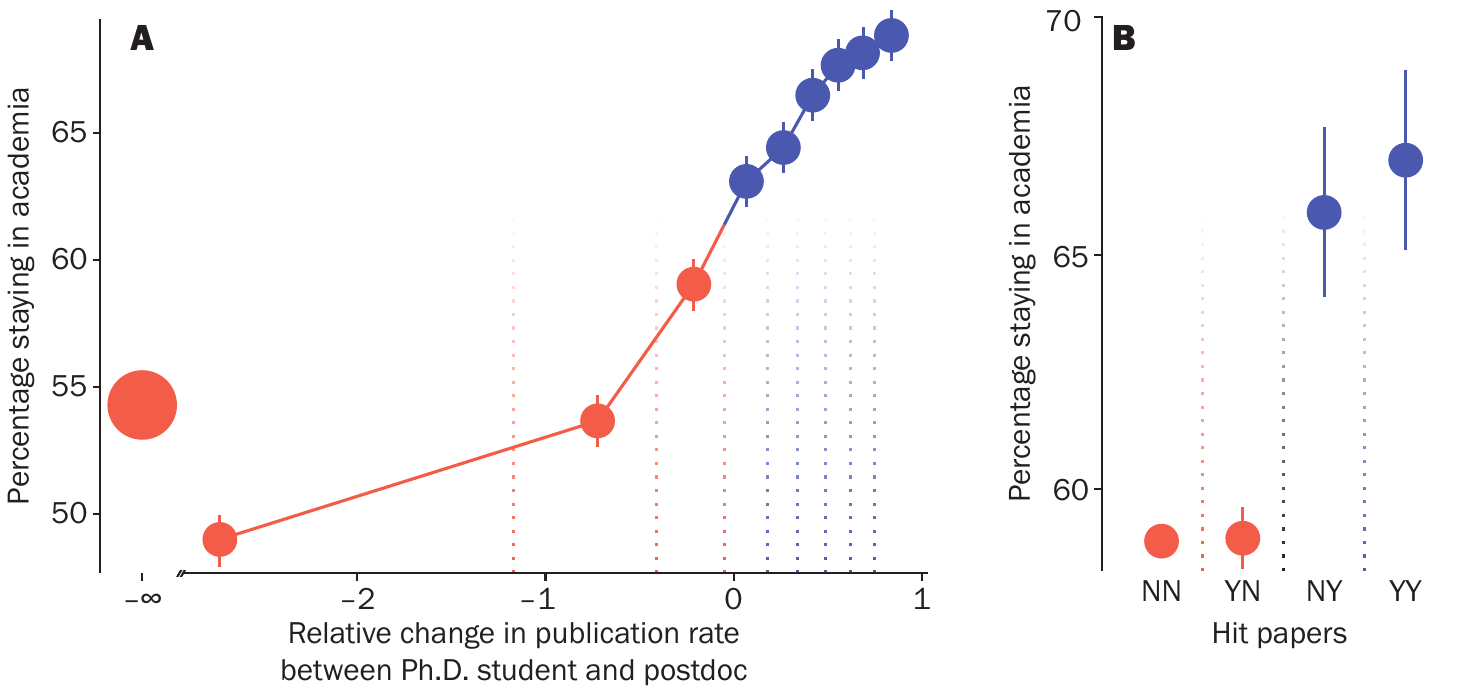}
\caption{\textbf{Who drops out of academia after a postdoc.} Panel A shows the percentage of postdocs proceeding to a faculty position as a function of the relative change in their publication rate between their doctorate studies and their postdoc. The bins divide the number of observations equally, except the $-\infty$ point (representing those without any publications as postdocs), which includes more observations as indicated by the larger area. Panel B shows the effects of having a hit paper during the Ph.D.\ program or the postdoc. The category ``YN'' means people with one or more hit papers during their Ph.D. training but none during their postdoc---Y(es) for PhD.\ student, N(o) for postdoc---and so on. The individuals without a hit paper as a postdoc are significantly less likely to pursue a faculty career than those with at least one (with an effect size given by Cohen's $d>12$). In both panels, the error bars represent standard errors.}
\label{fig:bottleneck}
\end{figure*}

In our (and other~\cite{eco_sci}) data, $41\%$ of the postdocs leave academia. This dropout represents a systemic bottleneck that is unrelated to the qualifications of the individuals involved---many postdocs are, by design, destined to leave academia. There is ample anecdotal evidence of the haphazard nature of postdoctoral training---promising new doctors who are handed projects that just happen not to work out~\cite{woolston2020postdoc,arnold2014stressed}. This suggests that comparing the publication rates between the periods as a graduate student and a postdoc could provide valuable insights. In Fig.~\ref{fig:bottleneck}A, we can indeed observe a strong correlation between the change in publication rate and the chance of becoming a faculty. A more detailed analysis shows that this is true independently of the absolute publication rate.

Productivity is, however, not the only academic capital. Quality, or impact, measured via citation counts is perhaps even more important for a career~\cite{way}. There is a prevailing idea that early citation success is a prerequisite for a faculty position~\cite{early}. This has prompted several authors to measure the career impact of \textit{hit papers}~\cite{early,fernandes_etal,network_effects}. In Fig.~\ref{fig:bottleneck}B, we show that the probability of staying in academia increases dramatically if they are able to achieve a \textit{hit paper}---a publication that finds itself in the top five percent of its field's year-end citation top list. Surprisingly, postdoc publications seem to matter more than those of Ph.D.\ students. Together with Fig.~\ref{fig:bottleneck}A, these results point to the importance of the postdoctoral period for the beginning of a faculty career.

\section*{Paths to success for early-year faculty}

We now leave the factors influencing whether there is an academic career on the other side of the postdoctoral bottleneck to investigate the direct effects of the postdoctoral training on one's career success. Quantifying success is a notorious problem. This is particularly clear in academia, made up of a mosaic of individuals with a multitude of career and life goals~\cite{sauermann2016pursue}. Instead of trying to capture the full complexity, we use standard academic performance indicators, capturing productivity and citation impact, as a starting point for a success measure tailor-made for early career faculty.

The direct effects of one's postdoctoral training should be the strongest soon after the postdoc. Thus, we focus on the scientific output between two and four years from the first faculty position. The first two-year gap is to clear out publications of research produced during the postdoc. The subsequent two-year measuring period was inspired by the definition of the journal impact factor. To characterize an individual's output, the most basic statistic would be to count the number of papers, their sum, or the maximum of acquired citations. Since both productivity and quality are awarded, we adapt the standard compromise---the h-index of the papers published in the period between two and four years after one's first faculty job. To avoid confusion with the common h-index, we refer to it as the $\eta$-index.

\begin{figure*}[t!]
\includegraphics[width=\linewidth]{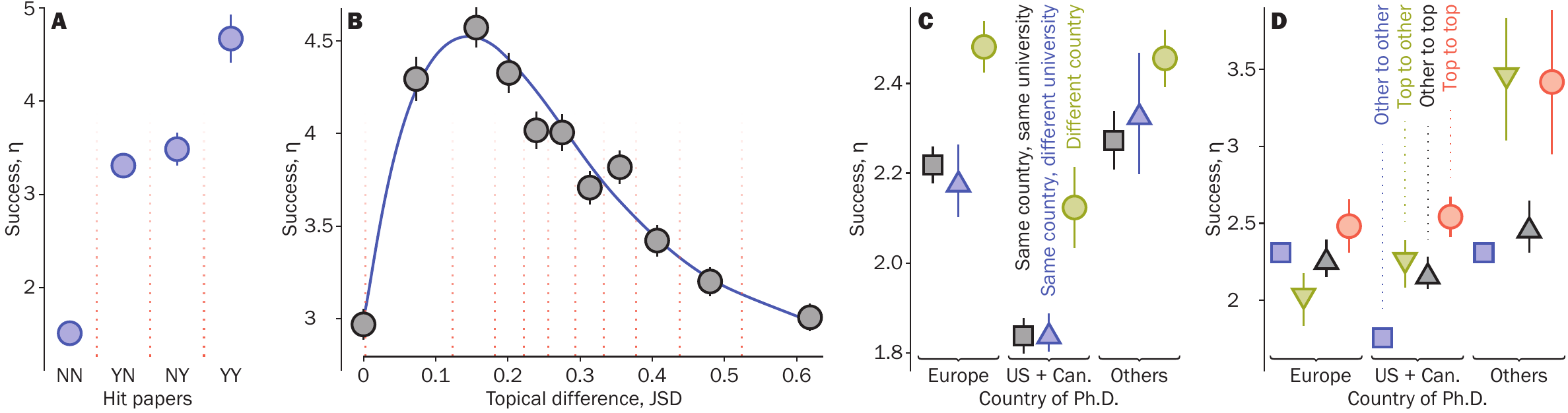}
\caption{\textbf{Factors influencing the success of early career scientists.} Panel A shows the average values of our success metric $\eta$ for the same categories as Fig.~\ref{fig:bottleneck}B (those in these categories that stayed in academia after their postdocs). Panel B displays $\eta$ as a function of the topical difference between the publications as a doctoral student and a postdoc measured by the Jensen-Shannon divergence (JSD) of the annotated disciplines of the publications. The smooth line is a fitted fourth-order polynomial. The vertical lines show the (equal sample number) bins (except for the zero divergence observations that form one bin). For reference, in our case (with 19 topical categories), changing from publishing on only one topic to publishing on only another gives a JSD of $0.83$. Panel C shows the average $\eta$ value for physical mobility---whether the Ph.D.\ moved to another country, another university in the same country, or stayed at the same university. Finally, panel D shows a similar plot for different classes of postdoctoral trajectories---whether the move to or from a top-10 university or research institution, or stays in the top-10 vs.\ other categories, and how these moves depend on the region of the Ph.D.\ granting institution---Europe, The US and Canada, the rest of the world.}
\label{fig:success}
\end{figure*}

In a first observation, the more ephemeral measure of publication success we employed above---whether one achieved a hit paper as a Ph.D.\ student or postdoc---not only predicts whether one stays in academia but also the early career success in terms of $\eta$ for those that do stay (see Fig.~\ref{fig:success}A). One difference is that for those with a hit paper during only one of the Ph.D.\ or postdoctoral periods, it does not matter during which of these periods the hit occurred. However, having hit papers during both these periods boosts $\eta$ significantly, and lacking hit papers correspondingly lowers $\eta$.

For the individual, the postdoc is an opportunity to ``gain scientific, technical, and professional skills that advance the professional career'' \cite{maxine_singer}. Gaining skills suggests widening one's repertoire in a different environment; Fig.~\ref {fig:success}B shows the future performance as a function of the topical change between the Ph.D.\ and postdoctoral publications as measured by the Jensen-Shannon divergence of the topics of publications (see Materials and Methods). Indeed, we can see that a moderate change of topic correlates with future success.

Changing the institution for the postdoc, and perhaps making it an international experience, is another way of broadening one's repertoire. In Fig.~\ref{fig:success}C, we compare the performance between different groups of postdoctors depending on whether they stay at their Ph.D.\ granting institution, move domestically, or move to another country. We see that moving abroad for a postdoc is associated with a higher $\eta$ ($d=0.14$), whereas for those staying in the same country, moving to another university has a negligible effect $d<0.01$.

Another type of mobility, more commonly studied in the literature, is in the space of prestige. Refs.~\cite{clauset_systematic,hanneman} found academia to be remarkably hierarchical, where new faculty hires are likely to have a Ph.D.\ from an institution higher in the hierarchy. For postdocs, the flow is reversed, with $2.4$ times more people moving to a top institution than away from one. The institutional prestige did not matter much for the $\eta$ score (Fig.~\ref{fig:success}D), except for young faculty members with a Ph.D.\ from outside of Europe, Canada, and the US, where having a Ph.D.\ from a prestigious institution seems influential ($d>0.40$).

\section*{Dicussion and conclusions}

We have found the postdoctoral period to be no less critical than the Ph.D.\ in determining future academic careers---those whose productivity went down during the postdoc, and those without a hit paper during this period are significantly more likely to drop out of academia than others. This sheds new light on the many papers reporting the importance of the Ph.D.-granting institution on future faculty hires~\cite{clauset_systematic,wapman_quantifying,network_effects,eun_dynamic}. Of course, a Ph.D.\ is an important factor in determining the location of the postdoc. Indeed, the causal diagrams determining one's career beyond the postdoc are complex, and omitting the postdoc seems like a fallacious choice.

Another insight is that while several papers describe the postdoc as a wait for a faculty job~\cite{i_r,s_ma}, this is poor advice for the postdocs themselves---those who are active and mobile fare better, even in their research as young faculty members. I.e., the commonly reported advantage of diverse teams~\cite{ethnic} seems true also for the individual---a diverse academic experience gives an advantage.

To summarize, our paper calls for models of the academic job market that give postdoctoral training its deserved attention. Our findings should also encourage doctoral students to take a moderate step out of their immediate academic surroundings for their postdoc.

\appendix

\section{Data}

The data acquisition, further described in the \textit{SI Appendix}, starts from the Microsoft Academic Graph (MAG)---a corpus of approximately 257 million academic publications, annotated with 317 million authors, across 19 top-major disciplines, and the citation network within the data~\cite{sinha2015overview, wang2019review}. We link individuals in the Microsoft Academic Graph (MAG) to career information from publicly available CVs on an online professional network. This approach allows us to identify scientists who have served as postdoctoral researchers at some point in their academic careers versus those who have not. After filtering out career paths other than Ph.D. $\rightarrow$ postdoc $\rightarrow$ a faculty position or a non-academic position, we ended up with a sample of 45,572 individuals.

\subsection*{Data ethics:} All data used in this study is publicly available. Furthermore, the research was reviewed by New York University Abu Dhabi IRB (HRPP--2023--239).

\begin{SCfigure*}[\sidecaptionrelwidth][t]\label{fig:ill}
\includegraphics[width=0.667\textwidth]{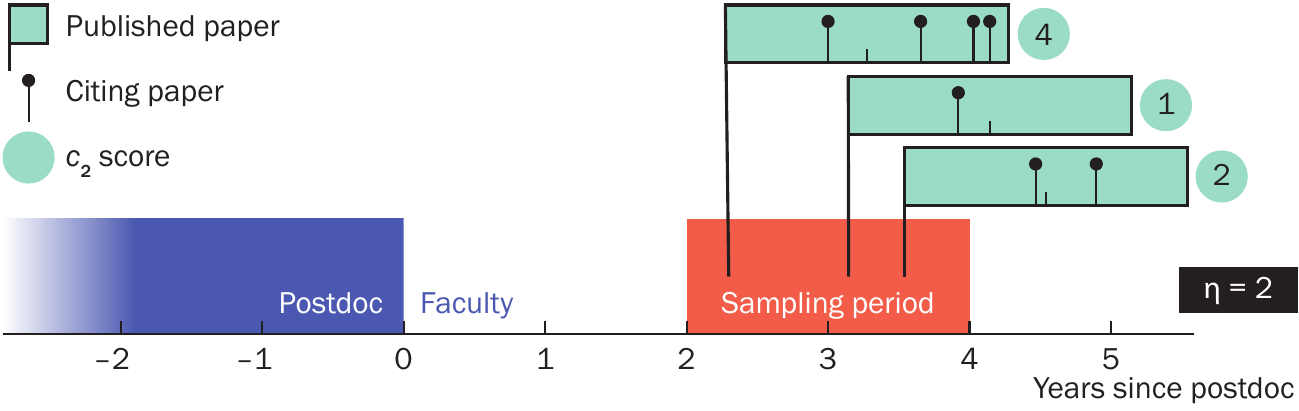}
\caption{\textbf{Illustration of the $\eta$ measure.} We focus on papers published between two and four years after the beginning of the faculty career. We gather the $c_2$ scores of these papers (the number of citations within two years of publication. Then, $\eta$ is the h-index of the set of $c_2$ scores.}
\end{SCfigure*}

\section{Measuring success}

\subsection{Hit papers}

As a measure of early success during the Ph.D. program and postdoctoral education, we follow Ref.~\cite{nearly} and measure whether someone had a \textit{hit paper} or not. Specifically, we rank the papers according to their amount of citations during each year within their disciplinary category (as annotated in MAG). Then, we categorize people based on whether a paper published during their Ph.D. studies (postdoctoral training) managed to become a hit paper in its discipline any year before the person's first faculty position. (We omit the information after the postdoc period to make the causal reasoning clearer.)

\subsection{The $\eta$-index}

Our measure of success for young faculty members builds on the ideas behind the h-index~\cite{h-index} and journal impact factor~\cite{jis} but covers the productivity of a new faculty member. We omit papers published within the first two years after the Ph.D. since they are not unlikely to have been a direct result of the postdoctoral training. We base our measure on the citations of the publications between two and four years after the postdoc. To extract one performance metric of the set of citation counts, we make use of the h-index~\cite{h-index}. The advantage of the h-index is that it is a very familiar compromise between productivity and citation impact of individual papers. Alternatives, like the raw citation or paper counts, also lead to the same conclusions, but somewhat less clearly.

More technically, let $c_2^i$ be the number of citations paper $i$ has acquired within the first two years of publication. Also assume an individual has $n$ papers with $c_2$ scores $(c_2^1,\dots,c_2^n)$ sorted in non-increasing order, then our success index is
\begin{equation}
    \eta = \max \left\{i : i < c_2^i\right\} .
\end{equation}
See Fig.~\ref{fig:ill} for an illustration. We chose the symbol $\eta$ since it corresponds to ``h'' in Greek.

\section{Measuring diversity}

Neither researchers nor papers need to be loyal to one discipline. To represent the disciplinary content of a publication, we use the indicator function of the annotated MAG categories. I.e., a binary vector $\mathbf{x}=(x_1,\dots,x_{19})$ where
\begin{equation}
    x_i = \left\{
    \begin{array}{ll} 1 & \text{if the paper belongs to category $i$}\\
    0 & \text{otherwise}
    \end{array} \right . .
\end{equation}
To represent the topical category of a publication list $a$, we simply sum the normalized $\mathbf{x}$-vectors.
\begin{equation}
\mathbf{p}(a)=\sum_j\frac{\mathbf{x}_j(a)}{|\mathbf{x}_j(a)|} ,
\end{equation}
where the sum is over all $a$'s publications and $|\:\cdot\:|$.

To measure the deviation between people at different periods of their careers---specifically, the Ph.D. studies versus their postdoctoral training---we use the Jensen-Shannon divergence.
\begin{align}
\text{JSD}(\mathbf{p}_{\text{Ph.D.}},\mathbf{p}_{\text{postdoc}}) = &H\left(\frac{\mathbf{p}_{\text{Ph.D.}}+\mathbf{p}_{\text{postdoc}}}{2}\right)\\ & - \frac{1}{2}\big (H(\mathbf{p}_{\text{Ph.D.}})+H(\mathbf{p}_{\text{postdoc}})\big ) , \nonumber
\end{align}
where $H(\cdot)$ denotes the Shannon entropy.

The advantage of the Jensen-Shannon divergence is that it has a solid, information-theoretical foundation and it is frequently used. The disadvantage, when applying it to our particular data, is that the contribution from every dimension is equal, meaning that even the most interdisciplinary individuals will be seemingly rather similar (since most elements of the $\mathbf{p}$-vectors will be zero). Another effect, desirable or not, is that the JSD also reflects differences in publication rates and the length of the publication lists. The shorter the lists are, the bigger the effects of fluctuations. This does affect Fig.~\ref{fig:success}B in the sense that less productive fields have, on average, lower $\eta$ and do contribute to both ends of the JSD spectrum. On one hand, one can (and we do) regard this as a feature rather than a bias. On the other hand, even if we divide the data into core disciplines or publication rates, one still obtains a peaked $\eta$-versus-JSD curve, in agreement with the conclusions of Fig.~\ref{fig:success}B.

\vspace{1cm}

\begin{acknowledgments}
YD was supported by the China Scholarship Council (CSC) program (No.\ 202106400071). QG was supported by Young Elite Scientist Sponsorship Program by BAST (Grant No.\
BYESS2023413). PH was supported by JSPS KAKENHI Grant Number JP 21H04595. We thank Hochan Lee for helping identify postdocs in MAG.
\end{acknowledgments}

\bibliographystyle{abbrv}
\bibliography{bib}

\end{document}